\documentclass[aps,reprint,showpacs,superscriptaddress]{revtex4-1}
\usepackage{epsf}
\usepackage{graphicx}
\usepackage{sidecap}
\usepackage{amsmath}

\usepackage{color}

\def \beq {\begin{equation}}
\def \eeq {\end{equation}}
\begin{document}

\title{Oscillatory surface dichroism of an insulating topological insulator Bi$_2$Te$_2$Se}

\author{M.~Neupane}
\affiliation {Joseph Henry Laboratory and Department of Physics,
Princeton University, Princeton, New Jersey 08544, USA}

\author{S.~Basak}
\affiliation {Department of Physics, Northeastern University,
Boston, Massachusetts 02115, USA}

\author{N.~Alidoust}\affiliation {Joseph Henry Laboratory and Department of Physics, Princeton University, Princeton, New Jersey 08544, USA}

\author{S.-Y.~Xu}
\affiliation {Joseph Henry Laboratory and Department of Physics, Princeton University, Princeton, New Jersey 08544, USA}


\author{Chang~Liu}
\affiliation {Joseph Henry Laboratory and Department of Physics,
Princeton University, Princeton, New Jersey 08544, USA}

\author{I.~Belopolski}\affiliation {Joseph Henry Laboratory and Department of Physics, Princeton University, Princeton, New Jersey 08544, USA}

\author{G. Bian}
\affiliation {Joseph Henry Laboratory and Department of Physics,
Princeton University, Princeton, New Jersey 08544, USA}

\author{J. Xiong}
\affiliation {Joseph Henry Laboratory and Department of Physics, Princeton University, Princeton, New Jersey 08544, USA}

\author{H.~Ji}\affiliation {Department of Chemistry, Princeton University, Princeton, New Jersey 08544, USA}

\author{S.~Jia}\affiliation {Department of Chemistry, Princeton University, Princeton, New Jersey 08544, USA}

\author{S.-K.~Mo}\affiliation {Advanced Light Source, Lawrence Berkeley National Laboratory, Berkeley, California 94305, USA}

\author{M. Bissen}\affiliation {Synchrotron Radiation Center, Stoughton, WI 53589-3097, USA}

\author{M. Severson}\affiliation {Synchrotron Radiation Center, Stoughton, WI 53589-3097, USA}



\author{H.~Lin}
\affiliation {Department of Physics, Northeastern University,
Boston, Massachusetts 02115, USA}

\author{N. P. Ong}
\affiliation {Joseph Henry Laboratory and Department of Physics, Princeton University, Princeton, New Jersey 08544, USA}

\author{T.~Durakiewicz}
\affiliation {Condensed Matter and Magnet Science Group, Los Alamos National Laboratory, Los Alamos, NM 87545, USA}

\author{R.~J.~Cava}
\affiliation {Department of Chemistry, Princeton University,
Princeton, New Jersey 08544, USA}

\author{A.~Bansil}
\affiliation {Department of Physics, Northeastern University,
Boston, Massachusetts 02115, USA}

\author{M.~Z.~Hasan}
\affiliation {Joseph Henry Laboratory and Department of Physics,
Princeton University, Princeton, New Jersey 08544, USA}

\date{\today}
\begin{abstract}

Using circular dichroism-angle resolved photoemission spectroscopy (CD-ARPES), we report a study of the effect of angular momentum transfer between polarized photons and topological surface states on the surface of highly bulk insulating topological insulator Bi$_2$Te$_2$Se. The photoelectron dichroism is found to be strongly modulated by the frequency of the helical photons including a dramatic sign-flip. Our results suggest that the observed dichroism and its sign-flip are consequences of strong coupling between the photon field and the spin-orbit nature of the Dirac modes on the surface. 
Our studies reveal the intrinsic dichroic behavior of topological surface states and point toward the potential utility of bulk insulating topological insulators in device applications.

\end{abstract}
\date{\today}
\maketitle

While the basic electronic structure and spin-momentum locking of topological insulators have been studied using surface sensitive probes such as angle-resolved photoemission spectroscopy and scanning tunneling microscopy \cite{Moore, Hasan, SCZhang, David Nature08, Matthew_BiSe, Chen_BiTe, David_PRL, Raghu, Hosur, spin Hall transistor, Ishida, Park,Park1,Gedik,Gedik_control, Gedik_review, Neupane, Suyang}, much remains to be discovered regarding their critical and strong response to light, electric or magnetic fields.
Such perturbations can selectively couple to different aspects of the surface wavefunction. The full wavefunction of the topological surface states (TSSs) is known to feature not only strong spin-orbit coupled texture but also its variation and modulation from layer to layer due to its finite penetration into the bulk \cite{Hugo_Dil, Park}. Therefore, it is of critical importance to understand the nature of electron-photon scattering process in the TSS. It is commonly believed that the single frequency dichroic signal reveals the spin (and/or orbital) texture of the material and also controls the photocurrent \cite{Park,Park1,Gedik,Gedik_control, Raghu, Hosur, spin Hall transistor}. However, in real materials, this apparently simple control process is further complicated by multiple factors including the presence of bulk bands at the Fermi level leading to surface bulk hybridization, quantum well formation and surface-bulk scattering thus masking the intrinsic response. Predictable control of the topological surface states has not yet been achieved.

In order to understand the intrinsic dichroic behavior of topological surface states it is important to study the effect of angular momentum transfer between the polarized photons and the surface states as a function of photon frequency and polarization in a highly bulk insulating topological insulator class where Fermi level lies within the bulk band-gap and cuts across the topological surface states only. We carried out circular dichroism-angle resolved photoemission (CD-ARPES) measurements on Bi$_2$Te$_2$Se (BTS221), a recently realized bulk resistive topological insulator (more than 6 $\Omega\cdot cm$). 
BTS221 sample shows much better insulating characteristics compared to Bi$_2$Te$_3$ or Bi$_2$Se$_3$, with an in--gap Fermi level, and is thus ideal for exploring the real origin of dichroic effects without complications related to interaction between the bulk and surface states. This is not possible in Bi$_2$Te$_3$ \cite{Oliver}.
We report that the intrinsic dichroism is strongly modulated by the frequency of photons including a dramatic sign-flip which further undergoes magnitude oscillations. Our results suggest a lack of unique experimental correspondence between the dichroism and spin-texture chirality (right or left handedness) for a specific photon frequency. We present theoretical calculations accounting for the Dirac-electron and helical-photon interaction and show that the sign-flip and the magnitude modulation in dichroism are consequences of the combined effect of strong coupling between the photon helicity and the spin-orbit texture of the massless Dirac modes and the projection of the multiple orbital-textures within the effective skin depth of the topological surface states.

Single crystalline samples of topological insulators were grown using the Bridgman method, which is detailed elsewhere \cite{BTS_Jia, Cava_BiTeS,BTS_Ando}. ARPES measurements for the low energy electronic structure were performed at the Synchrotron Radiation Center (SRC), Wisconsin, equipped with high efficiency VG-Scienta SES2002 electron analyzers, using the U9 VLS-PGM beam and the Advanced Light Source (ALS), California, using BL10 equipped with high efficiency  R4000 electron analyzers.
The polarization purity is better than 99\% for horizontal polarization (HP) and better than 80\% for right circularly polarized (RCP) and left circularly polarized (LCP) light.
Samples were cleaved {\it in situ} and measured at 20 K in a vacuum better than 1 $\times$ 10$^{-10}$ torr.
Energy and momentum resolution were better than 15 meV and 1\% of the surface Brillouin zone (BZ), respectively.
We theoretically calculate the CD response on the surface of BTS221, where the
electronic structure of BTS221 is modeled by the tight-binding theory with
the parameter fitted by the GGA results. The ARPES matrix element effects is considered in the electron-photon scattering process. (See \cite{SOM} for details).

\begin{figure*}
\centering
\includegraphics[width=15.0cm]{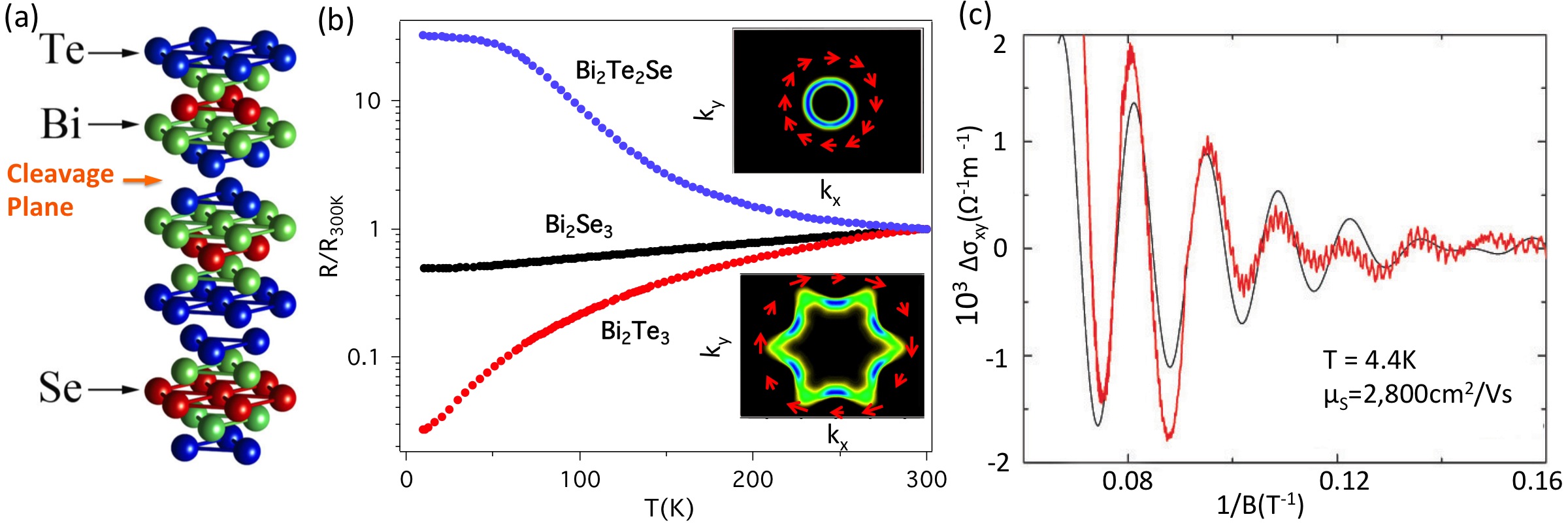}
\caption{(a) Crystal structure of BTS221. Red, green and blue circles represent the Se, Bi and Te atoms, respectively. 
(b) Normalized in-plane resistivities (R/R$_{300}$) plotted as a function of temperature (T) for BTS221. The resistivity profiles of  Bi$_2$Se$_3$ and Bi$_2$Te$_3$ are added for comparison.
Insets show the Fermi surface plots for BTS221 (upper panel) and Bi$_2$Te$_3$ (lower panel). Arrows around the FS represent the in-plane spin-texture.
 (c) Shubnikov-de Haas oscillation measurements on the topological surface of BTS221 (see \cite{BTS_sdh} for details).}
\end{figure*}

The crystalline symmetry, the cleavage plane (Te-layer) and sample characterization for BTS221 are shown in Fig. 1. It is believed that the reduction in the bulk conductivity is possible in BTS221 due to the confinement of Se atoms within the central layer which likely suppresses the Se vacancy generation as well as reduces the antisite defects between Bi and Te atoms. Comparative resistivity profiles show significant degree of bulk insulation in BTS221 with respect to prototype materials such as Bi$_2$Te$_3$ and Bi$_2$Se$_3$. Based on the period of oscillations in high-field transport, we obtain an averaged 2D carrier concentration $n_s \sim 1.7\times10^{12}$ cm$^{-2}$ and hence a Fermi momentum of $k_F \sim 0.047$
$\mathrm{\AA}^{-1}$. Applying a standard Dingle analysis to the SdH amplitudes, we infer a surface mobility $\mu_s=2,800$ cm$^2$/Vs and a Fermi velocity $v_F=6\times10^5$ m/s in our samples \cite{BTS_sdh}. The non-conducting behavior of the bulk and the in-gap Fermi level in our samples reduce the possibility of interaction of bulk and surface states, which is also evident from the high degree of surface state contribution to transport typically seen in the quantum oscillation data  \cite{BTS_sdh}. These results are in qualitative agreement with conventional bandstructure measurements of BTS221 \cite{Neupane, Suyang}. BTS221 samples thus provide an ideal platform to explore the intrinsic CD effect theoretically expected from the topological surface states, which is not possible with metallic Bi$_2$Te$_3$ TI \cite{Oliver}.

\begin{figure*}
\centering
\includegraphics[width=15.50cm]{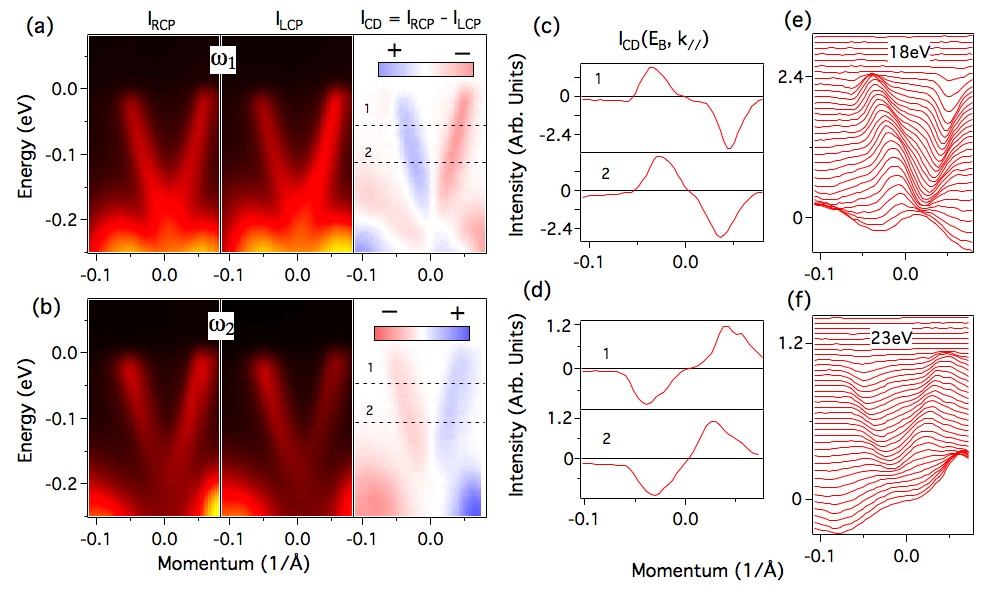}
\caption{(a) High-resolution ARPES measurements of Bi$_2$Te$_2$Se for right circularly polarized (RCP) light, left circularly polarized (LCP) light and the photoelectron circular dichroism ($I_{RCP} - I_{LCP}$) measured with photon energy $\omega_1$ =18 eV. (b) Analogus measurements  as in (a) for photon energy $\omega_2$ =23 eV. These spectra are measured along the $\bar\Gamma-\bar M$ high symmetry momentum-space cut. (c) The measured CD values for binding energies of 50 meV and 100 meV as marked on the $I_{CD}$ plot of (a) by black dashed lines and denoted by numbers 1 and 2. (d) Similar measurements as (c) for (b). (e) and (f) The momentum distribution curves of CD spectra for 18 eV and 23 eV photons, respectively. See \cite{SOM} for additional CD measurements data.}
\end{figure*}

The ARPES dispersion maps of surface states for BTS221 are shown in Fig. 2(a) while the experimental geometry used for the CD measurements is shown in the inset of Fig. 3(a). The sample surface is parallel to the XY plane and circularly polarized photons (spiral arrow) propagate in the XZ plane at an angle $(\theta)$ of 50$^{\circ}$ to the sample surface normal. The chemical potential is found to lie within the bulk band gap cutting across the TSS only. A nearly isotropic surface Fermi surface without any significant hexagonal deformation is seen which suggests that this system can be thought of as a material realization of a nearly-ideal Dirac Fermion gas near the native chemical potential. This also indicates a near-absence of interaction between bulk and surface states (in contrast to the hexagonally warped lower Dirac cone of Bi$_2$Te$_3$ shown in the inset of Fig. 1(b) \cite{Fu}). The Dirac node in BTS221 is found to be nearly buried within the bulk valence band, which makes the surface state in the lower cone degenerate with bulk bands. As a result, the intrinsic CD effect associated with the lower Dirac cone cannot be clearly disentangled from the bulk. We therefore focus on the detailed CD behavior of the upper Dirac band.

A clear surface state CD response on the photoelectron signal from the upper Dirac cone is observed where the +k Dirac branch is positive and the -k Dirac branch is negative in CD intensity (Fig. 2). The magnitude of the CD response signal defined as $I_{CD}$=($I_{RCP}-I_{LCP}$)/($I_{RCP}+I_{LCP}$) is observed to be about 20\% for incident photons with an energy of 18 eV in BTS221 for electrons with binding energy of about 150meV, well below the chemical potential. This CD behavior is qualitatively consistent with previous work on other TIs such as Cu$_x$Bi$_2$Se$_3$ \cite{Ishida} and Bi$_2$Se$_3$ \cite{Gedik, Park, Hoffman}.
Previously such CD response has been used to derive the details of spin-texture and chirality under the assumption that the response measured at a single frequency qualitatively samples the complete surface state wavefunction properties. In a multi-orbital system where surface state penetrates more than the very top layer a single photon energy may not capture the full details of the wavefunction. Indeed, analogously measured CD response but for the 23 eV photons shows, in our data Fig. 2(b), a momentum space reversal of CD sign per Dirac band, namely, the +k Dirac branch is negative whereas the -k Dirac branch is positive. The reversal of CD between 18 and 23 eV is also seen in our systematic measurements of the momentum distribution profiles (Figs. 2(c-f)). We further study the CD response of these samples with photons of intermediate energies to study the functional dependence on photon frequency or energy. Fig. 3 reveals the systematics of the CD response, magnitude and sign, in BTS221 within photon energies from 20 to 31 eV. We found that the reversal of CD between 18 eV and 23 eV is in fact a part of the full oscillation profile.

\begin{figure*}
\centering
\includegraphics[width=15.00cm]{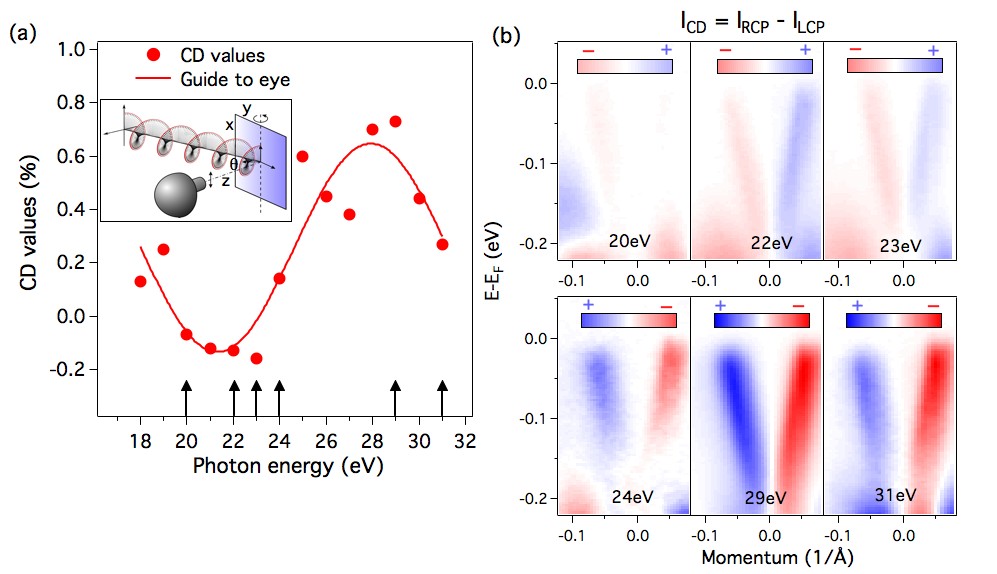}
\caption{(a) The measured CD values are plotted as a function of photon energy. The photoelectron CD value is estimated as  $I_{CD}$=($I^{+}-I^{-}$)/($I^{+}+I^{-}$)  for data taken with momentum k $\sim$ $-$0.05 $\AA^{-1}$ and binding energy $\sim$ 100 meV. Arrows represent the photon energies of representative CD-ARPES spectra presented in (b).
Inset shows the geometry of the ARPES measurement (see text for details). (b)  ARPES plots of the circular dichroic photoemission with various photon energies. The corresponding photon energies are noted on the plots. The spectra taken with 20-23 eV exhibit  negative CD values (top row) while it is taken with 24-31 eV exhibit positive CD values (bottom row). See \cite{SOM} for additional data.}
\end{figure*}

In order to further check the generality of our observation, we perform surface CD-ARPES measurements on two other TI systems, namely Bi$_{1.4}$Sb$_{0.6}$Te$_{1.5}$S$_{1.5}$ (BiSbTeS) and  the prototype Bi$_2$Te$_3$. (See \cite{SOM} for details). The flipping of the CD sign is also observed in these compounds. The flipping of CD in metallic Bi$_2$Te$_3$ is in agreement with ref. \cite{Oliver}. These systematic measurements imply that the CD modulation and sign-flip behavior are likely to be a general property of the topological surface states beyond BTS221.
Our results indicate that the intrinsic CD of the topological surface states is strongly modulated with photon energy and the existence of the sign-flip suggests that the CD signal can not be a straightforward reflection of the spin texture of the initial ground state \cite{Lanzara, Zhu, Oliver}. Taking the experimental data alone suggest a lack of unique correspondence between the dichroism and spin-texture chirality namely left or right handedness which is critical for the mirror Chern number measurement reflecting the class of topological order.

In order to understand the general trend observed in our experimental data, we carry out DFT(GGA) based first-principles calculations with consideration of the photoemission electron-photon scattering process and its microscopic relation to the measured CD profile. We write the photoemission matrix element as follows:
\begin{eqnarray}
  \label{M}
M_{\pm}({\bf k_f}) &=
& \sum_{i,nlm}\mathcal{J}_{i,nlm}(k_f)
 \Big( \pm i a\chi_{nlm,\downarrow(\uparrow)}
\notag\\
 +(1 \mp i b)\chi_{nlm}\Big)
\end{eqnarray}
where the `+' and `$-$' signs refer to the right and left circularly polarized light, respectively. Here $a$ and $b$ are constants related to the crystal potential, $\mathcal{J}_{i,nlm}(k_f) =  (-i)^l e^{i\bf{k_f}\cdot\bf{R_i}} F_{nl}(k_f)Y_{lm}(\theta_{k_f},\phi_{k_f})$ and $\chi_{nlm}=\chi_{nlm,\uparrow}-\chi_{nlm,\downarrow}$. $F_{nl}(k_f) =\int r^2drj_{l}(k_fr)R_{nl}(r)$ is the form factor associated with the atomic orbital ($nlm$), $j_l$ is the spherical Bessel function, ${\bf k}_f$ is the momentum of the free electron final state and $Y_{lm}$ is the spherical harmonic for the angular variables of ${\bf k}_f$. The initial state is expanded into atomic orbital ($nlm$) of the $i$-th atom in the unit cell at position $\bf{R_i}$, $|i \rangle = R_{nl}(r)Y_{lm}(\theta,\phi)|\chi \rangle_{nlm}$ where the spin function is $|\chi \rangle_{nlm} = \chi_{nlm,\uparrow} |\uparrow\rangle + \chi_{nlm,\downarrow} |\downarrow\rangle$, $|\uparrow\rangle$ and $|\downarrow\rangle$ being the spin eigenstates for the quantization axis along z-direction. 
 The circular dichroism is a consequence of combination of spectral function of the topological surface state and the photoemission matrix element under left and right circularly polarized light as expressed below:

 \begin{eqnarray}
  \label{Icd}
I_{CD}({\bf k}_f,{\bf k},E) \propto \Big(|M_{+}({\bf k}_f, \uparrow, \downarrow)|^2-
\notag\\
|M_{-}({\bf k}_f, \uparrow, \downarrow)|^2\Big)\mathcal{W}({\bf k},E)
\end{eqnarray}
where $\mathcal{W}({\bf k},E)$ is the spectral function of the initial topological surface states, and ${\bf k}$, $E$,  $\uparrow$, and $\downarrow$ are the momentum, the energy, and the spin of the initial topological surface states, respectively.
The detail of calculation is given in \cite{SOM}. It can be seen from equation (2) that the term of $(|M_{+}|^2- |M_{-}|^2)$  is a function of the spin of  initial ground states of the topological surface states. More importantly, we also note that photoelectron momentum {\bf k$_f$} is a photon energy-dependent quantity, and consequently $I_{CD}$ is also a function of incident photon energy, as follows from equation (2) . Thus our model naturally explains the experimental observation of the CD signal dependence on incident photon energy.
In Figs. 4(a) and 4(b) we plot the calculated CD spectra for two sets of photon energies and the sign-flip oscillation is seen in Fig. 4(c).
The structure factor $e^{i\bf{k_f}\cdot\bf{R_i}}$ also indicates that the surface states may have a spatially dependent orbital mixing and it can be probed by varying the perpendicular component of the Bloch wave vector, $k_{f\perp}$, through its dependence on photon energy. As a result, the CD of the surface states could have a non-trivial dependence on the $k_{f\perp}$ of the emitted electron. Thus the photon energy dependent photoelectron interference effect plays a role in shaping up the spectra by masking the chirality of the spin-texture of the initial electronic states.

\begin{SCfigure*}
\centering
\includegraphics[width=11.0cm]{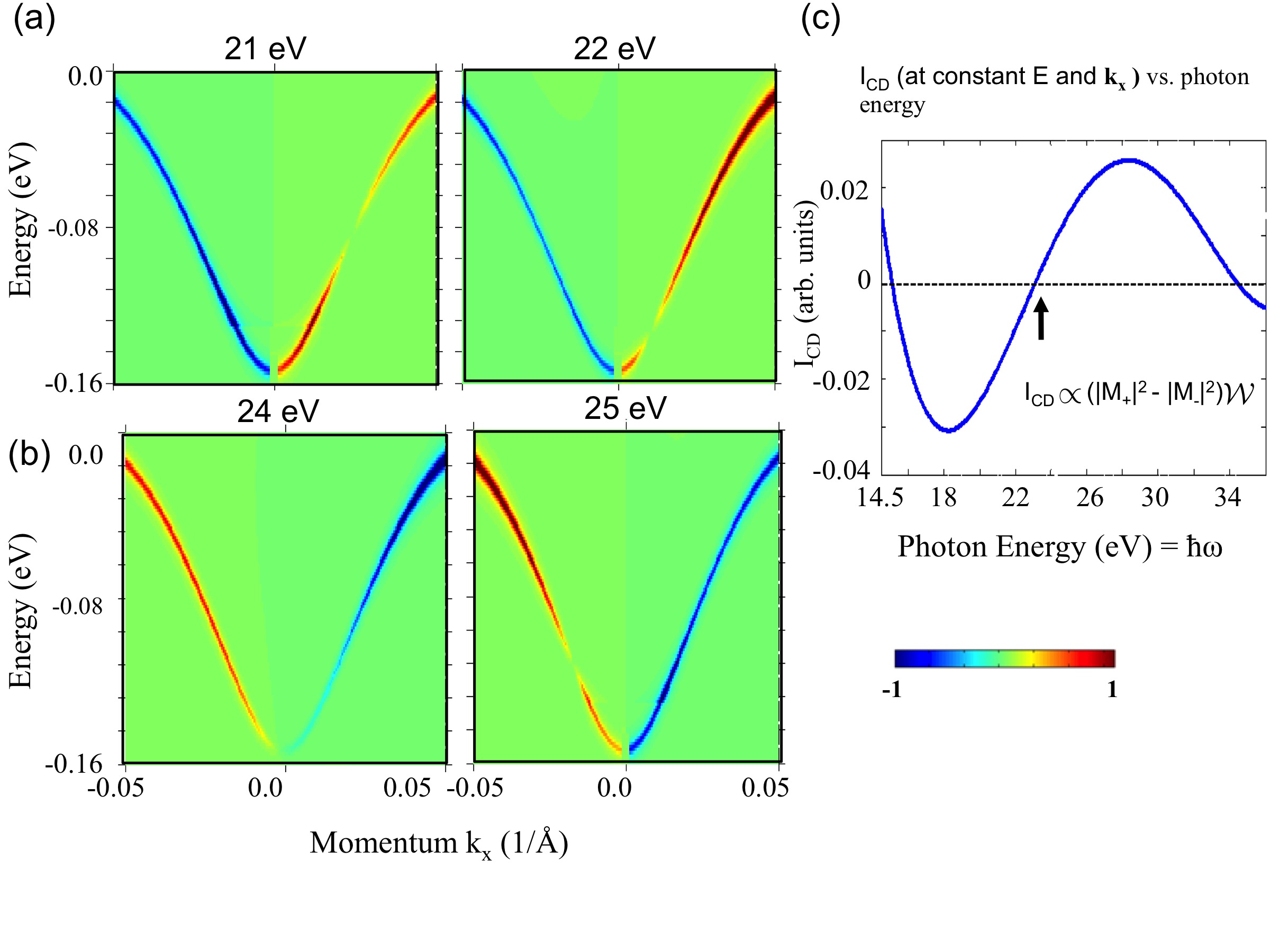}
\caption{Model calculations are carried out for photon energies of (a) 21 eV and 22 eV, and (b) 24 eV and 25 eV. The change in the sign of CD can be observed by comparing  spectra shown in (a) and (b). (c) The theoretically calculated $I_{CD}$ at constant $k_x$ and $E$ as a function of photon energy. The black arrow indicates the photon energy  value where flipping of CD sign is expected in our theoretical model for BTS221.} 
\end{SCfigure*}

According to the helical spin-texture, the CD of the lower Dirac cone is expected to have opposite sign with respect to that of the upper Dirac cone. However, CD of the lower Dirac cone is found to have a complex profile in which at certain photon energies it even shows the same sign as the upper Dirac cone. This behavior of CD of the lower Dirac cone can possibly be attributed to the intermixing between the bulk valence bands and surface states well below the Fermi level.
Furthermore, we use the three-step model of photoemission to determine the ARPES matrix element which is technically very different from the one-step model \cite{Oliver}. Despite the use of the one-step formulation of ARPES in treating the photoemission process as a single coherent event, it is often difficult to adduce physical microscopic insight into the origin of spectral features resulting from these quite complex calculations. The three-step modeling, on  the other hand, is very transparent, allowing straightforward disentanglement of various factors at play in the observed  spectral features by considering the  microscopic processes at play. Thus, our model is able to identify the photon energy dependent structure factor $e^{i\bf{k_f}\cdot\bf{R_i}}$ and $\mathcal{J}_{i,nlm}(k_f)$ to be the physical quantities controlling the sign-flip of the CD profile. This model can be  applied further to disentangle the contributions of various orbitals as a function of photon energy. 

In conclusion, we presented systematic photon energy dependent circular polarization response of photoelectrons in ARPES revealing an anomalous behavior of the CD signal on the topological insulator surfaces.
Our experimental results supported by our theoretical calculations suggest that measured CD response not only depends on the orbital/spin angular momentum of the initial states but also on the photon energy sampled, mixed-orbital content and the details of the coupling mechanism of initial state to the electric field of the incident light.
Our experimental findings reveal a rich response behavior of topological surface states thus open new avenues in understanding and controlling topological insulator properties with polarized light.

\textit{Acknowledgements.} This work was primarily supported by DARPA Meso grant $\#$ N66001-11-1-4110. The use of Synchrotron Radiation Center (SRC) was supported by NSF DMR-0537588 under the external user agreement.
We acknowledge Peter Riseborough and Luca Perfetti for discussions. T.D. at LANL acknowledges support from Department of Energy, Office of Basic Energy Sciences, Division of Material Sciences. M.Z.H. acknowledges Visiting Scientist support from LBNL and additional support from DOE/BES and the A. P. Sloan Foundation.

\bigskip

\*Correspondence and requests for materials should be addressed to M.Z.H. (Email: mzhasan@princeton.edu).

\end{document}